\documentclass[12pt,a4paper]{article}
\usepackage{amsfonts,amssymb,amsthm}
\usepackage{cite}
\usepackage[T2A]{fontenc}
\usepackage[cp1251]{inputenc}
\usepackage[english]{babel}
\usepackage{amsmath}%
\usepackage[unicode]{hyperref}
\usepackage[left=1.5cm,right=1.5cm,
    top=2cm,bottom=3cm,bindingoffset=0cm]{geometry}

\def\const {\hbox{const}}

\allowdisplaybreaks
\numberwithin{equation}{section}

\title{\bf Miura type transformations \\ for integrable lattices in 3D}
\author{\bf I. T. Habibullin, A. R. Khakimova and A. U. Sakieva}
\date{}
\begin{document}
\maketitle
\begin{abstract} The article studies a class of integrable semidiscrete equations with one continuous and two discrete independent variables. Miura type transformations are obtained that relate the equations of the class. A new integrable chain of this type is found, for which the Lax pair is presented. Integrable in the sense of Darboux reductions of the chain are discussed, for small order reductions complete sets of integrals are constructed. Continuum limits for the chain are discussed. A method for finding particular solutions of chains based on integrable in sense of Darboux reductions is proposed. The effectiveness of the method is illustrated by an  example.
\end{abstract}

\maketitle

\section{Introduction}

Nonlinear integrable differential-difference equations with one continuous and two discrete variables are intermediate objects between the class of integrable nonlinear equations of dimension 1 + 2, such as the Toda chain, the Kadomtsev-Petviashvili equation, the Davey-Stewartson equation, etc. (see \cite{BKon}, \cite{KonS}), and the class of fundamental discrete Hirota-Miwa type equations (see \cite{Hirota1}, \cite{Hirota2}, \cite{Miwa}). They also have applications in the theory of Laplace transformations in the class of linear differential-difference equations of hyperbolic type\cite{AdS1999}. More precisely, the sequence of Laplace invariants satisfies some integrable differential-difference equation of the form \eqref{one}, related by a point transformation to equation 7) from the list below. Such chains can find application in studying the nonlinear dynamics of localized magnetic inhomogeneities in such magnetically ordered substances as ferromagnets, antiferromagnets with weak ferromagnetism, magnetoelastic and magnetoelectric interactions \cite{ShamsutdinovHab2009}, \cite{Shamsutdinov2009}.
 Equations of the type under consideration have potential applications in the problem of describing dislocations in media with a microstructure, as well as in the nonlinear theory of elastic and nonelastic deformations accompanied with deep reconstruction of an
initially ideal lattice: switching of interatomic bonds, changing of the class of symmetry, formation of new phases,
singular defects and heterogeneities, fragmentation of the lattice (see, for instance, \cite{Kunin75}, \cite{AeroBulygin}).

All known integrable differential-difference equations with one continuous and two discrete independent variables can be reduced to the following form
\begin{equation}\label{one}
u^j_{n+1,x}=f(u^j_{n,x},u^{j+1}_n, u^j_n, u^j_{n+1}, u^{j-1}_{n+1}),
\end{equation}
where the sought function $u=u^j_n(x)$ depends on the continuous $x$ and discrete $n$ and $j$. The subscript $x$ denotes the derivative $u_x=\frac{d}{dx}u$.
Below we give a list of currently known examples of integrable equations of the form \eqref{one}:
\begin{enumerate}
	\item[1)]
$ u^j_{n+1,x}=u^j_{n,x}+e^{u^j_{n+1}-u^{j+1}_n}-e^{u^{j-1}_{n+1}-u^j_n},$
\item[2)]
$ u^j_{n+1,x}=u^j_{n,x}+e^{u^{j-1}_{n+1}-u^j_{n+1}}-e^{u^{j-1}_{n+1}-u^j_n}-e^{u^j_n-u^{j+1}_n}+e^{u^j_{n+1}-u^{j+1}_n},$
\item[3)]
$ u^j_{n+1,x}=u^j_{n,x}\frac{{(u^j_{n+1})^2}}{{u^{j-1}_{n+1}}u^{j+1}_n},$
\item[4)]
$ u^j_{n+1,x}=u^j_{n,x}\frac{u^j_{n+1}-u^{j+1}_n}{u^{j-1}_{n+1}-u^j_n}, $
\item[5)]
$ u^j_{n+1,x}=u^j_{n,x}\frac{u^j_{n+1}(u^j_{n+1}-u^{j+1}_n)}{u^{j+1}_n(u^{j-1}_{n+1}-u^j_n)}, $
\item[6)]
$ u^j_{n+1,x}=u^j_{n,x}\frac{(u^{j-1}_{n+1}-u^j_{n+1})(u^j_{n+1}-u^{j+1}_n)}{(u^{j-1}_{n+1}-u^j_n)(u^j_n-u^{j+1}_n)}, $
\item[7)]
$ u^j_{n+1,x}=u^j_{n,x}+e^{{-u^{j+1}_n+u^j_n}+u^j_{n+1}-u^{j-1}_{n+1}}, $
\item[8)]
$ u^j_{n+1,x}=u^j_{n,x}-e^{u^{j+1}_n}+e^{u^j_n}+e^{u^j_{n+1}}-e^{u^{j-1}_{n+1}}, $
\item[9)]
$ u^j_{n+1,x}=u^j_{n,x}+(u^j_{n+1}-u^j_n)(u^{j+1}_n-u^j_{n+1}-u^j_n+u^{j-1}_{n+1}). $
\end{enumerate}
Equation 1) was derived in \cite{Hab95} on the basis of iterations of the B\"acklund transformation for the two-dimensional Toda lattice. Equations 7) and 8) were found in \cite{AdS1999} while studying the Laplace invariants of linear differential-difference equations of hyperbolic type with variable coefficients. Equations 2)-6) were obtained in \cite{FNR} by an original method based on a preliminary study of the dispersionless limits of chains for integrability, followed by a return transition to the class of chains \eqref{one} as a result of dispersion deformations. Equation 9) is new: it is obtained in the present work.

A distinctive feature of nonlinear chains with three independent variables is that they allow reductions in the form of systems of differential-difference equations with complete sets of characteristic integrals. Such systems of equations are called Darboux integrable. As a rule it is possible to find in an explicit form general solution to Darboux integrable systems at least for the case of small orders. This circumstance allows one finding localized particular solutions for the lattices in 3D via the Darboux integrable reductions.  In our articles \cite{Hab13}, \cite{HP2017}, \cite{HabKh21}, \cite{KuzHabKh23}, an algorithm for classifying integrable equations of the form \eqref{one} is developed, where the presence of a hierarchy of the Darboux integrable reductions is taken as a classification criterion.

One can check that the equations from the list above are not related to each other by point changes of variables. However, the equations could be transformed from one to the other using more complex irreversible Miura type transformations, which would be useful in finding explicit solutions to the equations (see, for instance, \cite{Nijhoff84}, \cite{ShYam97}, \cite{Levi}). Therefore, the task of searching for such transformations is in demand. The study of this problem is the aim of the present work.

The results of working with Miura type transformations for the class of lattices \eqref{one} exceeded our expectations. They led to the discovery of a new integrable chain 9).  This equation is unique in some extent, since it has a polynomial (quadratic) right hand side, while the right hand sides of the other chains are either rational or exponential. 

Let us briefly discuss on the content of the article. In \S2 we suggest an algorithm for searching differential or/and difference substitutions generated by local conservation laws. In \S 3 we apply the algorithm to equations 1)--9). We have established connections between the equations. In \S 4 we give the Lax pair for the chain 9) and calculate for it the continuum limits of two types. We also consider examples of the small order Darboux integrable reductions admitted by the chain. For the reduced systems the complete sets of the characteristic integrals are constructed. An explicit particular solution to chain 9) is presented obtained via the reduced equation. 

\section{Local conservation laws and Miura type transformations.}

To find the Miura transformation that converts an equation of a given class \eqref{one} into some other equation of the same class, we use the following scheme. Let's first represent the equation as a local conservation law of the form:
\begin{equation}\label{loc1}
(D_n-1)A=(D_j-1)B
\end{equation}
or the form
\begin{equation}\label{loc2}
D_xA=(D_j-1)B, 
\end{equation}
where $A$ and $B$ are functions depending on the sought function $u=u^j_n(x)$, as well as on its derivatives with respect to $x$ and shifts with respect to discrete variables $j$ and $n$. Here $D_n$ and $D_j$ denote operators for shifting a function argument in $n$ and $j$, for example, $D_ny(n)=y(n+1),\quad D_jy(j)=y(j+1)$, and $D_x$ stands for the operator of the total differentiation with respect to $x$. If the representation \eqref{loc1} takes place, we define a new function $v=v^j_n(x)$ by introducing a potential, i.e. assuming
\begin{equation}\label{pot}
(D_j-1)v^j_n=A.
\end{equation}
Then substituting instead of the function $A$ its representation \eqref{pot} into relation \eqref{loc1} we arrive at the equality
\begin{equation}\label{next}
(D_j-1)(D_n-1)v^j_n=(D_j-1)B,
\end{equation}
from which, applying the operator inverse  to $(D_j-1)$, we find another important relation connecting the variables $u^j_n$ and $v^j_n$:\begin{equation}\label{usl}
(D_n-1)v^j_n=B+C.
\end{equation}
Since we do concentrate only on autonomous transformations we assume that $C$ is a constant. In the examples considered below, it is usually possible to completely exclude the variable $u^j_n$ from the pair of relations \eqref{pot}, \eqref{usl}. In such a case, we arrive at some equation for the new function $v^j_n$.

The second way to look for the desired substitution uses representation \eqref{loc2}. Here we introduce a new function ${\omega}^j_n$ due to the relation:
\begin{equation}\label{usl1}
D_x{\omega}^j_n=B.
\end{equation}
 Now the equation yields a relation
\begin{equation}\label{usl2}
D_xA=D_x\left(D_j-1\right){\omega}^j_n.
\end{equation}
Integrating the latter, we find, in addition to \eqref{usl1}, the following constraint that the variable ${\omega}^j_n$ should satisfy,
\begin{equation}\label{usl3}
\left(D_j-1\right){\omega}^j_n=A+\const.
\end{equation}
To obtain the lattice equation for ${\omega}^j_n$ it is necessary to get rid the variable $u^j_n$ by means of the relations \eqref{usl1} and \eqref{usl3}.

As one can see from the examples below, this approach very often leads to success. In particular, it is in this way that we managed to find Miura type transformations for each of the equations from the list under consideration.

\section{Searching for the transformations}

In this section we apply the algorithm discussed above to the lattices 1)-9). It is interesting to note that the set of equations falls into two non-overlapping subclasses. Most of the equations belong to the first subclass, namely 1), 3), 4), 7), 8), 9). Only three equations fall into the second class: 2),~5),~6).

\subsection{Connection between equations 1) and 9)}

Obviously equation 1) can be rewritten as follows
\begin{equation}\label{Toda}
(D_n-1)u^j_{n,x}=(D_j-1)e^{u^{j-1}_{n+1}-u^j_n}.
\end{equation}
According to the rule above we introduce a new variable $v^j_n$ by setting
\begin{equation}\label{T2}
u^j_{n,x}=v^{j+1}_n-v^j_n=(D_j-1)v^j_n,
\end{equation}
then formulas \eqref{Toda}, \eqref{T2} imply   
\begin{equation*}
(D_j-1)(D_n-1)v^j_n=(D_j-1)e^{u^{j-1}_{n+1}-u^j_n}
\end{equation*}
or, the same
\begin{equation}\label{T3}
(D_n-1)v^j_n=e^{u^{j-1}_{n+1}-u^j_n}+C.
\end{equation}
Let us differentiate \eqref{T3} with respect to $x$ and replace derivatives $u^{j-1}_{n+1,x}$, $u^j_{n,x}$ due to \eqref{T2}. As a result we arrive at the equation 
\begin{equation}\label{new}
v^j_{n+1,x}-v^j_{n,x}=(v^j_{n+1}-v^{j-1}_{n+1}-v^{j+1}_{n}+v^j_n)(v^j_{n+1}-v^j_n-C),
\end{equation}
that is reduced to 9) by a point transformation $x=-\bar{x}$, $v^j_n=u^j_n-Cn$.

\subsection{Connection between equations 2) and 5)}
Now we derive a substitution generated by the equation 5). To this end we first simplify 5) by the change of variables $u^j_n=e^{v^j_n}$:
\begin{equation*}
v^j_{n+1,x}=v^j_{n,x}\frac{e^{v^j_{n+1}-v^{j+1}_n}-1}{e^{v^{j-1}_{n+1}-v^j_n}-1}.
\end{equation*}
Then we represent the obtained equation in the form of the local conservation law:
\begin{equation}\label{loc21}
(D_n-1)\ln {v^j_{n,x}}=(D_j-1)\ln \left(e^{v^{j-1}_{n+1}-v^j_n}-1\right).
\end{equation}
Let us introduce a new dependent variable ${\omega}^j_n$ as
\begin{equation}\label{zam}
\left(D_j-1\right){\omega}^j_n=\ln v^j_{n,x}  
\end{equation}
or
\begin{equation*}
v^j_{n,x}=e^{{\omega}^{j+1}_n-{\omega}^j_n}.
\end{equation*}
Then from \eqref{loc21}, \eqref{zam} we obtain the equation
\begin{equation*}
\left(D_j-1\right)\left(D_n-1\right){\omega}^j_n=\left(D_j-1\right)\ln \left(e^{v^{j-1}_{n+1}-v^j_n}-1\right),
\end{equation*}
which implies:
\begin{equation}\label{3**}
\left(D_n-1\right){\omega}^j_n=\ln \left(e^{v^{j-1}_{n+1}-v^j_n}-1\right)+C.
\end{equation}
One can easily verify that the variable $v^j_n$ is completely eliminated from formulas \eqref{zam}, \eqref{3**} and their consequences in a rather simple way. Indeed, by differentiating \eqref{3**} we get 
\begin{equation}\label{4**}
{\omega}^j_{n+1,x}-{\omega}^j_{n,x}=\frac{e^{v^{j-1}_{n+1}-v^j_n}\left(v^{j-1}_{n+1,x}-v^j_{n,x}\right)}{e^{v^{j+1}_{n+1}-v^j_n}-1}.
\end{equation}
The right hand side of the formula is greatly simplified due to equalities \eqref{zam}, \eqref{3**} and as a result the equation reduces to the form
\begin{equation}\label{5**}
{\omega}^j_{n+1,x}-{\omega}^j_{n,x}=\left(1+e^{{\omega}^j_n-{\omega}^j_{n+1}+C}\right)\left(e^{{\omega}^j_{n+1}-{\omega}^{j-1}_{n+1}}-e^{{\omega}^{j+1}_n-{\omega}^j_n}\right).
\end{equation}
Up to the linear transformation ${\omega}^j_n=-u^j_n+Cj$ equation \eqref{5**} coincides with equation 2).

\subsection{Connection between equations 3) and 1)}
Let us turn to equation 3). Taking the logarithm of both sides of the equation we get an identity
\begin{equation}\label{ur300}
\ln u^j_{n+1,x}-\ln u^j_{n,x}=2\ln u^j_{n+1}-\ln u^{j-1}_{n+1}-\ln u^{j+1}_n,
\end{equation}
that implies a local conservation law of the form
\begin{equation}\label{ur30}
\left(D_n-1\right)\ln \frac{u^j_{n,x}}{u^j_n}=\left(D_j-1\right)\ln \frac{u^{j-1}_{n+1}}{u^j_n}.
\end{equation}
Having introduced a new variable
 $h^j_n$ by setting
\begin{equation}\label{ur31}
\left(D_j-1\right) h^j_n=\ln \frac{u^j_{n,x}}{u^j_n},
\end{equation}
we deduce from \eqref{ur30} the equality
\begin{equation}\label{z4}
\left(D_j-1\right)\left(D_n-1\right)h^j_n=\left(D_j-1\right)\ln\frac{u^{j-1}_{n+1}}{u^j_n},
\end{equation}
that gives rise to
\begin{equation}\label{z5}
\left(D_n-1\right)h^j_n=\ln\frac{u^{j-1}_{n+1}}{u^j_n}+C,
\end{equation}
where $C$ is assumed to be constant. Now we differentiate \eqref{z5} respect to $x$ and find  
\begin{equation*}
h^j_{n+1,x}-h^j_{n,x}=\frac{u^{j-1}_{n+1,x}}{u^{j-1}_{n+1}}-\frac{u^j_{n,x}}{u^j_n},
\end{equation*}
which is easily simplified due to \eqref{ur31}. As a result we obtain the lattice 
\begin{equation*}
h^j_{n+1,x}-h^j_{n,x}=e^{h^j_{n+1}-h^{j-1}_{n+1}}-e^{h^{j+1}_n-h^j_n}.
\end{equation*}

The latter coincides with the discrete Toda lattice 1) up to the linear change of the independent discrete variables $j=-\bar{j}, n=\bar{j}+\bar{n}$.

\subsection{Connection between equations 3) and 7)}
We note that for equation 3) one can derive one more substitution. To this end we use the equation
\eqref{ur300}  rewritten as
\begin{equation}\label{ur34}
\left(D_n-1\right)\ln {u^j_{n,x}}=\left(2D_n-D_nD^{-1}_j-D_j\right)\ln u^j_n.
\end{equation}
Assuming that $\ln u^j_n=\left(D_n-1\right)v^j_n$, we obtain 
\begin{equation}\label{ur35}
u^j_n=e^{v^j_{n+1}-v^j_n}.
\end{equation}
Then combining \eqref{ur34} and \eqref{ur35} we arrive at
\begin{equation}\label{ur36}
u^j_{n,x}=e^{2v^j_{n+1}-v^{j-1}_{n+1}-v^{j+1}_n}
\end{equation}
and by differentiating \eqref{ur35} we find a similar relation
\begin{equation}\label{ur37}
u^j_{n,x}=\left(v^j_{n+1,x}-v^j_{n,x}\right)e^{v^j_{n+1}-v^j_{n}}.
\end{equation}
Comparing the last two representations for $u^j_{n,x}$ one can get the equation
\begin{equation}\label{ur38}
v^j_{n+1,x}-v^j_{n,x}=e^{-v^{j+1}_n+v^j_n+v^j_{n+1}-v^{j-1}_{n+1}},
\end{equation}
which is nothing but equation 7).

\subsection{Connection between equations 4) and 1)}

Now we study equation 4) in the list. It can easily be represented as a local concervation law of the form
\begin{equation}\label{ur41}
\left(D_n-1\right)\ln u^j_{n,x}=\left(D_j-1\right)\ln \left(u^{j-1}_{n+1}-u^j_n\right).
\end{equation}
We introduce a new unknown $s^j_n$ by taking
\begin{equation}\label{ur42}
\left(D_j-1\right)\ln s^j_n=\ln u^j_{n,x}.
\end{equation}
Then from \eqref{ur41} it follows that
\begin{equation*}
\left(D_j-1\right)\left(D_n-1\right)\ln s^j_n=\left(D_j-1\right)\ln \left(u^{j-1}_{n+1}-u^j_n\right).
\end{equation*}
By applying to the latter equation the operator $\left(D_j-1\right)^{-1}$ we obtain the second part of the searched substitution:
\begin{equation}\label{ur43}
\left(D_n-1\right)\ln s^j_n=\ln \left(u^{j-1}_{n+1}-u^j_n\right)+\ln C, \quad C=\const.
\end{equation}
Let's simplify \eqref{ur43} a little, after which, applying the operator $D_x$ to both its sides, we obtain a nonlinear lattice
\begin{equation*}
\frac{s^j_{n+1,x}}{s^j_n}-\frac{s^j_{n+1}s^j_{n,x}}{\left(s^j_n\right)^2}=C\left(\frac{s^j_{n+1}}{s^{j-1}_{n+1}}-\frac{s^{j+1}_n}{s^j_n}\right),
\end{equation*}
that is reduced to the discrete Toda lattice 1) by the point transformation
 $s^j_n=e^{-u^j_n},\quad x\mapsto{Cx}$.

\subsection{Connection between equations 6) and 2)}

Let us look for the Miura type transformations generated by equation 6). By taking logarithm of both sides of 6) we get a relation
\begin{equation*}
\left(D_n-1\right)\ln u^j_{n,x}=\left(D_n-D_j\right)\ln \left(u^{j-1}_n-u^j_n\right)+\left(D_j-1\right)\ln \left(u^{j-1}_{n+1}-u^j_n\right).
\end{equation*}
Afterwards by using the decomposition $D_n-D_j=D_n-1-\left(D_j-1\right)$ one can rewrite the relation as a local conservation law of the form:
\begin{equation}\label{ur61}
\left(D_n-1\right)\ln \frac{u^j_{n,x}}{u^{j-1}_n-u^j_n}=\left(D_j-1\right)\ln \frac{u^{j-1}_{n+1}-u^j_n}{u^{j-1}_n-u^j_n}.
\end{equation}
Then we introduce a new variable $h^j_n$ by the rule
\begin{equation}\label{ur62}
\left(D_j-1\right)\ln h^j_n=\ln \frac{u^j_{n,x}}{u^{j-1}_n-u^j_n}.
\end{equation}
Applying the reasoning above to \eqref{ur61}, \eqref{ur62} one can derive:
\begin{equation}\label{ur63}
\left(D_n-1\right)\ln h^j_n=\ln \frac{u^{j-1}_{n+1}-u^j_n}{u^{j-1}_n-u^j_n}+\ln C.
\end{equation}
Let us rewrite Miura transformation \eqref{ur62}, \eqref{ur63} in a more convenient form
\begin{equation}\label{ur64}
\frac{u^j_{n,x}}{u^{j-1}_n-u^j_n}=\frac{h^{j+1}_n}{h^j_n},\qquad C\frac{u^{j-1}_{n+1}-u^j_n}{u^{j-1}_n-u^j_n}=\frac{h^j_{n+1}}{h^j_n},
\end{equation}
which implies that
\begin{equation}\label{ur65}
\frac{1}{C}\frac{u^j_{n,x}}{u^{j-1}_{n+1}-u^j_n}=\frac{h^{j+1}_n}{h^j_{n+1}}.
\end{equation}
Differentiating \eqref{ur63} we find a relation
\begin{equation}\label{diff}
\frac{h^j_{n+1,x}}{h^j_{n+1}}-\frac{h^j_{n,x}}{h^j_n}=\frac{u^{j-1}_{n+1,x}-u^j_{n,x}}{u^{j-1}_{n+1}-u^j_n}-\frac{u^{j-1}_{n,x}-u^j_{n,x}}{u^{j-1}_n-u^j_n}=
\frac{u^{j-1}_{n+1,x}}{u^{j-1}_{n+1}-u^j_n}-C\frac{h^{j+1}_n}{h^j_{n+1}}-\frac{u^{j-1}_{n,x}}{u^{j-1}_{n}-u^j_n}+\frac{h^{j+1}_n}{h^j_n}.
\end{equation}
We simplify some of the terms in \eqref{diff} due to the lattice 6) and get:
\begin{align*}
\frac{u^{j-1}_{n+1,x}}{u^{j-1}_{n+1}-u^j_n}-\frac{u^{j-1}_{n,x}}{u^{j-1}_n-u^j_n}&=\frac{u^{j-1}_{n,x}}{u^{j-1}_n-u^j_n}\left(\frac{u^{j-2}_{n+1}-u^{j-1}_{n+1}}{u^{j-2}_{n+1}-u^{j-1}_n}-1\right)=\\
&=C\frac{h^j_n}{h^{j-1}_{n+1}}\frac{u^{j-1}_n-u^{j-1}_{n+1}}{u^{j-1}_n-u^j_n}=C\frac{h^j_n}{h^{j-1}_{n+1}}\left(1-\frac{u^{j-1}_{n+1}-u^j_n}{u^{j-1}_n-u^j_n}\right)=C\frac{h^j_n}{h^{j-1}_{n+1}}\left(1-\frac{1}{C}\frac{h^j_{n+1}}{h^j_n}\right).
\end{align*}
Finally we obtain a lattice
\begin{equation*}
\frac{h^j_{n+1,x}}{h^j_{n+1}}-\frac{h^j_{n,x}}{h^j_n}=\frac{h^{j+1}_n}{h^j_n}-C\frac{h^{j+1}_n}{h^j_{n+1}}+C\frac{h^j_n}{h^{j-1}_{n+1}}-\frac{h^j_{n+1}}{h^{j-1}_{n+1}},
\end{equation*}
which is related to the lattice 2) by a point transformation
 $h^j_n=C^ne^{-p^j_n}$:
\begin{equation*}
p^j_{n+1,x}-p^j_{n,x}=e^{p^{j-1}_{n+1}-p^j_{n+1}}-e^{p^{j-1}_{n+1}-p^j_n}-e^{p^j_n-p^{j+1}_n}+e^{p^j_{n+1}-p^{j+1}_n}.
\end{equation*}

\subsection{Connection between equations 8) and 1)}

Let us look for a substitution that would connect lattices 8) and 1). At first we rewrite equation 8) in the form of a local conservation law:
\begin{equation}\label{ur71}
\left(D_n-1\right)u^j_{n,x}=\left(D_j-1\right)\left(e^{u^{j-1}_{n+1}}-e^{u^j_n}\right).
\end{equation}
Supposing that 
\begin{equation}\label{3.24'}
 u^j_{n,x}=\left(D_j-1\right)p^j_{n,x},
\end{equation}
we introduce a new dependent variable $p^j_n$. Then  obviously it follows from \eqref{ur71}, \eqref{3.24'} that
\begin{equation}\label{ur72}
\left(D_j-1\right)\left(D_n-1\right)p^j_{n,x}=\left(D_j-1\right)\left(e^{u^{j-1}_{n+1}}-e^{u^j_n}\right).
\end{equation} 
From the latter one can derive the second part of the Miura type transformation
\begin{equation}\label{ur73}
\left(D_n-1\right)p^j_{n,x}=e^{u^{j-1}_{n+1}}-e^{u^j_n}+C_1.
\end{equation}
The expression $u^j_n=\left(D_j-1\right)p^j_n+C_2$ is an evident consequence of \eqref{3.24'}, therefore we have
\begin{equation}\label{3.27}
e^{u^j_n}=e^{C_2}e^{p^{j+1}_n-p^j_n}.
\end{equation}
Let us eliminate the variables $u^j_n, u^{j-1}_{n+1}$ from \eqref{ur73} due to \eqref{3.27} and obtain a lattice of the form:
\begin{equation*}
p^j_{n+1,x}-p^j_{n,x}=e^{C_2}\left(e^{p^j_{n+1}-p^{j-1}_{n+1}}-e^{p^{j+1}_n-p^j_n}\right)+C_1.
\end{equation*}
Changing the variables by setting  $p^j_n=q^j_n+C_1xn$ we arrive at the lattice
\begin{equation*}
q^j_{n+1,x}-q^j_{n,x}=e^{C_2}\left(e^{q^j_{n+1}-q^{j-1}_{n+1}}-e^{q^{j+1}_n-q^j_n}\right),
\end{equation*}
that converts to the discrete Toda lattice 1) under replacement
 $j=-\bar{j}, n=\bar{j}+\bar{n}, x=e^{-C_2}\bar{x}$. 

\section{On a novel integrable chain of the form (1.1)}

In this section, we focus on a detailed study of the found chain 9): we represent the Lax pair, calculate the continuum limit, and study Darboux-integrable reductions. For reduced systems of small dimensions, characteristic integrals are found in both directions $n$ and $x$. We also briefly discuss on the method of constructing particular solutions of the chain 9) by means of its reduction.

Let us begin with the Lax pair. It is checked straightforwardly that system of linear equations 
\begin{equation}\label{Lax9}
\begin{aligned}
&{\psi}^j_{n+1}=\left(v^j_n-v^j_{n+1}\right){\psi}^{j+1}_n+{\psi}^j_{n}, \\
&{\psi}^j_{n,x}=\left(v^j_n-v^{j-1}_n\right){\psi}^j_n-{\psi}^{j-1}_n
\end{aligned}
\end{equation}
is compatible if and only if function $v=v_n^j(x)$ is a solution to the lattice 9).

\subsection{Computation of the continuum limit for the lattice 9)}

In this section, we will calculate the continuum limit of the chain 9) under unbounded refinement of the difference grid in both directions $n$ and $j$. At first we perform 
a scaling transformation $y=\delta^2x$ in the chain \eqref{new}, and represent it as follows:
\begin{equation}
\delta^2\frac{d}{dy}\left(v^j_{n+1}-v^j_n\right)=\left(v^j_{n+1}-v^j_n\right)\left[\left(v^{j+1}_n-2v^j_n+v^{j-1}_n\right)+\left(v^{j-1}_{n+1}-v^{j-1}_n\right)-\left(v^j_{n+1}-v^j_n\right)\right]. \label{lattice}
\end{equation}
Let us assume that the sought function $v^j_n(x)$ is of the form $v^j_n(x)=u(t,z,y)$, where $t=\delta^2n$, $z={\delta}j$ and $\delta$ is a small positive parameter. 
Now we estimate the difference derivatives with respect to the variables $n$ and $j$ appearing in \eqref{lattice} by virtue of the Taylor formula 
\begin{equation}
v^j_{n+1}-v^j_n=u(t+\delta^2,z,y)-u(t,z,y)=\delta^2u_t(t,z,y)+O(\delta^4), \qquad \delta\mapsto0,
\end{equation}
\begin{align*}
v^{j-1}_{n+1}-v^{j-1}_n&=u(t+\delta^2,z-\delta,y)-u(t,z-\delta,y)=\\
&=u(t,z,y)+\delta^2u_t(t,z,y)-{\delta}u_z(t,z,y)+\frac{1}{2}\delta^2u_{zz}(t,z,y)-\\
&-u(t,z,y)+\delta u_z(t,z,y)-\frac{1}{2}\delta^2u_{zz}(t,z,y)+...+O(\delta^3)=\\
&=\delta^2 u_t(t,z,y)+O(\delta^3), \qquad \delta\mapsto0,
\end{align*}
\begin{align*}
v^{j+1}_n-2v^j_n+v^{j-1}_n&=u(t,z+\delta,y)-2u(t,z,y)+u(t,z-\delta,y)=\\
&=u(t,z,y)+{\delta}u_z(t,z,y)+\frac{1}{2}{\delta}^2u_{zz}(t,z,y)-2u(t,z,y)+\\
&+u(t,z,y)-{\delta}u_z(t,z,y)+\frac{1}{2}{\delta}^2u_{zz}(z,t,y)+...=\\
&={\delta}^2u_{zz}(t,z,y)+O(\delta^3), \qquad \delta\mapsto0.
\end{align*}
Now we substitute the found asymptotic representations of finite differences into the equation \eqref{lattice} and pass to the limit at $\delta\mapsto0$. As a result, we obtain the following dispersionless partial differential equation:
\begin{equation}
u_{ty}=u_tu_{zz}, \label{dispersionless}
\end{equation}
which is integrable as well \cite{FHKN}.

\subsection{Intermediate continuum limit}

Now we study the continuum limit of chain 9) assuming that sought function $u^j_n(x)$ depends on a small parameter $\varepsilon>0$ as follows $u^j_n(x)={\omega}^j(x,t)$ with $t=\varepsilon{n}$. Under the condition $\varepsilon\mapsto0$ lattice 9) tends to an equation of the Toda chain type \cite{ShYam97}
\begin{equation}\label{w1}
\omega^j_{xt}+\omega^j_t\left(\omega^{j+1}-2\omega^j+\omega^{j-1}\right)=0
\end{equation}
having the Lax pair of the form 
\begin{align}\label{w2}
\begin{aligned}
&\varphi^j_t=-\omega^j_t\varphi^{j+1}+\varphi^j,\\
&\varphi^j_x=\left(\omega^j-\omega^{j-1}\right)\varphi^j-\varphi^{j-1}
\end{aligned}
\end{align}
that is easily obtained from \eqref{Lax9} under the relation $\psi^j_n(x)=\varphi^j(x,t),$ when $\varepsilon\mapsto0$. A remarkable fact is that chains 9) and \eqref{w1} have a close connection. Indeed, by comparing the Lax pairs one can conclude that chain 9) realizes a sequence of the B\"acklund transformations for lattice \eqref{w1}. More precisely any solution $\omega^j$ of \eqref{w1} is converted by the B\"acklund transformation 
\begin{equation}\label{w3}
{\bar{\omega}}^j_x-\omega^j_x=\left({\bar{\omega}}^j-\omega^j\right)\left(-\omega^{j+1}+\omega^j+{\bar{\omega}}^j-{\bar{\omega}}^{j-1}\right)
\end{equation}
to another solution ${\bar{\omega}}^j$ of the lattice. We note that eigenfunctions $\varphi^j$ and ${\bar{\varphi}}^j$ of the system \eqref{w2} and respectively of the system
\begin{equation*}
{\bar{\varphi}}^j_t=-{\bar{\omega}}^j_t{\bar{\varphi}}^{j+1}+{\bar{\varphi}}^j, \quad {\bar{\varphi}}^j_x=\left({\bar{\omega}}^j-{\bar{\omega}}^{j-1}\right){\bar{\varphi}}^j-{\bar{\varphi}}^{j-1}
\end{equation*}
are connected with one another as follows
\begin{equation}\label{w4}
{\bar{\varphi}}^j=\left(\omega^j-{\bar{\omega}}^j\right)\varphi^{j+1}+\varphi^j.
\end{equation}

\subsection{Darboux integrable reductions of the lattice 9)}
Chain 9) admits the following cutting off boundary condition
\begin{equation}\label{421}
{v^j_n(x) |}_{j=j_0}=C,\quad C=\const
\end{equation}
preserving the integrability property. Imposing this kind of the conditions on two points $j=0$ and $j=N$ we obtain a finite field reduction of lattice 9):
\begin{align}\label{422}
\begin{aligned}
&v^{-1}_n=C_0,\\
&v^j_{n+1,x}-v^j_{n,x}=\left(v^j_{n+1}-v^j_n\right)\left(-v^{j+1}_n+v^j_n+v^j_{n+1}-v^{j-1}_{n+1}\right), \qquad 0\leq{j}\leq{N-1},\\
&v^N_n=C_1,
\end{aligned}
\end{align}
admitting the Lax representation of the form
\begin{align}
&\begin{cases}\label{sys1}
{\psi}^j_{n+1}=\left(v^j_n-v^j_{n+1}\right){\psi}^{j+1}_n+{\psi}^j_n, \qquad {0}\leq{j}\leq{N-1},\\
{\psi^N_{n+1}}={\psi}^N_n,
\end{cases}\\
&\begin{cases}\label{sys2}
{\psi}^0_{n,x}=\left(v^0_n-C_0\right){\psi}^0_n,\\
{\psi}^j_{n,x}=\left(v^j_n-v^{j-1}_n\right){\psi}^j_n-{\psi}^{j-1}_n, \qquad {1}\leq{j}\leq{N}
\end{cases}
\end{align}
deduced from \eqref{Lax9} under the constraints ${\psi}^{-1}_n={\psi}^{N+1}_n=0$, $v^{-1}_n=C_0$, $v^N_n=C_1$.
Let us concentrate on the finite field system \eqref{422}. It is worth noting that \eqref{422} admits complete sets of integrals in both directions $n$ and $x$. Recall that a function $I$ depending of the dynamical variables is an $n$-integral of the differential-difference system if it satisfies the relation $D_nI=I$ by virtue of the system. Integrals of the form $I=I(x)$ and $J=J(n)$ depending on $x$ and $n$ only are called trivial. We are only interested in non-trivial integrals. Let us give examples of integrals.

{\bf Example 1}. By setting $N=1$ in \eqref{422} we obtain a scalar equation for $v:=v^0$
\begin{equation}\label{*} 
v_{n+1,x}-v_{n,x}=\left(v_{n+1}-v_n\right)\left(v_{n+1}+v_n-C_0-C_1\right),
\end{equation}
which admits $n$-integral 
\begin{equation}\label{*1}
I=v_x-(v-C_0)^2-(v-C_1)(C_0-C_1)
\end{equation}
and $x$-integral
\begin{equation}\label{xint-1}
J=\frac{\left(v_n-v_{n-2}\right)\left(v_{n+1}-v_{n+3}\right)}{\left(v_n-v_{n+3}\right)\left(v_{n+1}-v_{n+2}\right)}.
\end{equation}

{\bf Example 2}. For the choice $N=2$ we arrive at the system for $u_n=v^0_n$, $v_n=v^1_n$
\begin{align}\label{sys-2}
\begin{aligned}
&u_{n+1,x}-u_{n,x}=\left(u_{n+1}-u_n\right)\left(-v_n+u_n+u_{n+1}-C_0\right),\\
&v_{n+1,x}-v_{n,x}=\left(v_{n+1}-v_n\right)\left(-C_1+v_n+v_{n+1}-u_{n+1}\right).
\end{aligned}
\end{align} 
The obtained system admits a complete set of $n$-integrals, i.e. a set of two independent $n$-integrals 
\begin{align*}
&I_1=u_{n,x}+v_{n,x}-\left(u_n-C_0\right)^2-\left(u_n-C_1\right)\left(C_0-v_n\right)-\left(C_1-v_n\right)^2,\\
&I_2=v_{n,xx}-2\left(v_n-C_1\right)v_{n,x}-\left(C_0-v_n\right)u_{n,x}-\left(v_n-C_0\right)\left(u_n-C_1\right)\left(u_n-v_n-C_0+C_1\right)
\end{align*}
and a complete set of $x$-integrals
\begin{align*}
&J_1=\frac{\left(\tau^{0}_{n}\tau^{1}_{n}+\tau^{0}_{n+1}\tau^{1}_{n+1}+\tau^{0}_{n}\tau^{1}_{n+1}\right)(\tau^{0}_{n+1}\tau^{1}_{n+1}+\tau^{0}_{n+2}\tau^{1}_{n+2}+\tau^{0}_{n+1}\tau^{1}_{n+2})}{\tau^{0}_{n}\tau^{1}_{n+2}\left(\tau^{0}_{n+1}\tau^{1}_{n}+\tau^{0}_{n+2}\tau^{1}_{n+1}+\tau^{0}_{n+2}\tau^{1}_{n}\right)},\\
&J_2=\frac{\tau^{0}_{n+3}\tau^{1}_{n}\left(\tau^{0}_{n+1}\tau^{1}_{n+1}+\tau^{0}_{n+2}\tau^{1}_{n+2}+\tau^{0}_{n+1}\tau^{1}_{n+2}\right)}{\tau^{0}_{n+2}\tau^{1}_{n+1}\left(\tau^{0}_{n+1}\tau^{1}_{n}+\tau^{0}_{n+2}\tau^{1}_{n+1}+\tau^{0}_{n+2}\tau^{1}_{n}+\tau^{0}_{n+3}\tau^{1}_{n+2}+\tau^{0}_{n+3}\tau^{1}_{n+1}+\tau^{0}_{n+3}\tau^{1}_{n}\right)},
\end{align*}
where $\tau^{0}_{n}=u_{n+1}-u_n$, $\tau^{1}_{n}=v_{n+1}-v_n$ (about complete set of integrals see, for instance, \cite{ZhiberK21}, \cite{HabKh22}).  

It can be proved that system \eqref{422} admits complete sets of $x$-integrals and $n$-integrals for any choice of $N$. For $N\geq3$ $x$-integrals have extremely large expression. They are found by the method of characteristic Lie-Rinehart algebras (see \cite{HabKh21}). As can be seen from the above examples, the $n$-integrals of the reduced system are much simpler. Note that the complete set of $n$-integrals to the system \eqref{422} for an arbitrary natural $N\geq1$ can be efficiently derived using the algorithm proposed in \cite[Lemma 2]{HabKh21}. An alternative approach to the problem of looking for integrals is proposed in \cite{Smirnov15}.

\subsection{Explicit solutions to the chain 9)}

The presence of complete sets of integrals in both characteristic directions $x$ and $n$ makes it possible to completely separate the variables. Indeed, according to the definition, each solution $v=(v^0, v^1,\dots,v^{N-1})$ of system \eqref{422} simultaneously satisfies two systems of equations, namely the system of ordinary differential equations 
\begin{equation}\label{ord}
I_1={\varphi}_1(x), \quad I_2={\varphi}_2(x), \quad ..., \quad I_N={\varphi}_N(x)
\end{equation}
and the system of ordinary discrete equations
\begin{equation}\label{ord1}
J_1={\psi}_1(n),\quad J_2={\psi}_2(n),\quad...,\quad J_N={\psi}_N(n),
\end{equation} 
where ${\varphi}_1(x), {\varphi}_2(x),...,{\varphi}_N(x)$ are arbitrary functions of $x$ and similarly ${\psi}_1(n), {\psi}_2(n),...,{\psi}_N(n)$ are arbitrary functions of the variable $n$.

In this case  common solution $\left\{v^j_n(x)\right\}^{N-1}_{j=0}$ of the  systems provides a solution to the system \eqref{422}.  We remark that solution of the reduced system \eqref{422} obviously can be prolonged to a particular solution of the original lattice 9).  However  for arbitrary $N$ system \eqref{ord} is rather difficult to solve. Here we discuss only a comparatively simple  case  $N=1$.
Let us derive a formula describing general solution to the lattice \eqref{*}. Due to the formula $DI=I$  one can conclude  that $I={\varphi}(x)$ i.e. function $v_n(x)$ is a solution to the equation
\begin{equation}\label{r1}
v_{n,x}-\left(v_n-C_0\right)^2-{\varepsilon}\left(v_n-C_1\right)={\varphi}(x), \quad \varepsilon=C_0-C_1.
\end{equation}
Since  $\varphi(x)$ is arbitrary  we can assume  that $\varphi(x)$ is represented as  
$$\varphi(x)=z_x-(z-C_0)^2-\varepsilon(z-C_1)$$ 
for some  function $z=z(x)$ that now is  considered as a new functional parameters. Let us  change the variables in \eqref{r1}  by taking $v_n=z+s$ and get a new equation  with unknown $s$:
\begin{equation}\label{r2}
s_x-s^2-s\left(\varepsilon+2\left(z-C_0\right)\right)=0,
\end{equation}
which has a form of the Bernoulli equation. Let us replace  $s=\frac{1}{g}$ and arrive  at a linear equation
\begin{equation}\label{r3}
g_x=-g\left(\varepsilon+2\left(z-C_0\right)\right)-1.
\end{equation}
We first solve the corresponding homogeneous equation 
$$g_x=-g\left(\varepsilon+2\left(z-C_0\right)\right).$$ 
We pass from the  functional parameter $z=z(x)$ to a new parameter $p(x)$ in such a way: 
$$p_x=\varepsilon+2(z-C_0).$$ 
Then the homogeneous equation 
$$g_x=-gp'$$ 
is easily integrated $g=Ce^{-p}$. Now supposing $C=C(x)$ we return to the equation \eqref{r3} and find  $C_x=-e^p$. We change the parameter in the equation again by setting $e^p={\omega}_x$, where $\omega$  is a new  parameter. As a result we get $C(x)=-\omega+K(n)$, where $K(n)$ is a constant of integration that might depend on $n$. Therefore we have  
$$g=\frac{K(n)-\omega}{{\omega}_x}\quad \mbox{so that} \quad s=\frac{{\omega}_x}{K(n)-\omega}.$$ 
To write down the final form of $v_n(x)$ we need in expression of $z$ in terms of $\omega$. To this end we exclude parameter  $p(x)$ from the relations 
$$z=C_0+\frac{1}{2}(p_x-\varepsilon), \quad e^p={\omega}_x$$ and find 
$$z=\frac{1}{2}\frac{{\omega}_{xx}}{{\omega}_x}+\frac{C_0+C_1}{2}.$$ 
Now we substitute  the found expressions into $v_n=s+z$ and obtain general solution  to the lattice~\eqref{*}:
\begin{equation}\label{r4}
v_n(x)=\frac{1}{2}\frac{{\omega}_{xx}}{{\omega}_x}+\frac{{\omega}_x}{K(n)-\omega}+\frac{C_0+C_1}{2},
\end{equation} 
where $\omega=\omega(x)$ is an arbitrary function of $x$, $K(n)$ is an arbitrary function of $n$.
Due to the fact  that the termination conditions  $v^{-1}_n=C_0$, $v^1_n=C_1$ are fully consistent with dynamics, by virtue of the original three-dimensional chain 9) the solution of the  reduced system \eqref{*} extends in an obvious way to the solution of the chain 9):
\begin{align*}
v^j_n(x)=\begin{cases}
C_0  &\mbox{for} \quad j\leq-1, \\
\displaystyle{\frac{{\omega}_{xx}}{2{\omega}_x}+\frac{{\omega}_x}{K(n)-\omega}+\frac{C_0-C_1}{2}} &\mbox{for} \quad j=0, \\
C_1 &\mbox{for} \quad j\geq1.\end{cases}
\end{align*}

\section*{Conclusions}

In the article integrable chains with three independent variables $x$, $n$, $j$ are considered. A list of currently known examples is given. Differential and difference substitutions linking equations of the list with each other are presented. A new integrable chain is found that is investigated in more detail. For this model the Lax pair is presented, the continuum limits are computed. Degenerate cutoff condition is pointed out which reduces the chain to a Darboux integrable system of differential-difference equations when it is imposed on two different points $(x,n,j=0)$ and $(x,n,j=N)$. For the reductions of small orders characteristic integrals are found. A method for constructing explicit particular solutions to the chain via reductions is proposed.


\begin{thebibliography}{5}

\bibitem{BKon} L. V. Bogdanov, B. G. Konopelchenko, “Analytic-bilinear approach to integrable hierarchies. I. Generalized KP hierarchy”, J. Math. Phys., \textbf{39} (1996), 4683--4700; “Analytic-bilinear approach to integrable hierarchies. II. Multicomponent KP and 2D Toda lattice hierarchies”, J. Math. Phys., \textbf{39} (1997), 4701--4728.

\bibitem{KonS} B. G. Konopelchenko, W. K. Schief, “Menelaus' theorem, Clifford configurations and inversive geometry of the Schwarzian KP hierarchy”, J. Phys. A: Math. Gen., \textbf{35}:29 (2002), 6125--6144.

\bibitem{Hirota1} R. Hirota, “Nonlinear partial difference equations”, II. Discrete-time Toda equations,
J. Phys. Soc. Japan, \textbf{43}:6 (1977), 2074--2078.

\bibitem{Hirota2} R. Hirota, “Discrete analogue of a generalized Toda equation”, J. Phys. Soc. Japan, \textbf{50}:11
(1981), 3785--3791.

\bibitem{Miwa} T. Miwa, “On Hirota’s difference equation”, Proc. Japan Acad. Ser. A, \textbf{58}:1 (1982), 9--12.

\bibitem{Nijhoff84} F. W. Nijhoff, H. W. Capel, G. L. Wiersma, G. R. W. Quispel, “B\"acklund transformations and three-dimensional lattice equations”, Physics Letters A, \textbf{105}:6 (1984), 267--272.

\bibitem{ShYam97} A. B. Shabat, R. I. Yamilov, “To a transformation theory of two-dimensional integrable systems”, Phys. Lett., A, 227:1–2 (1997), 15–23 , Elsevier (North-Holland), Amsterdam. 

\bibitem{Levi} D. Levi, M. Petrera, C. Scimiterna, R. Yamilov, “On Miura transformations and Volterra-type equations associated with the Adler-Bobenko-Suris equations”, SIGMA. Symmetry, Integrability and Geometry: Methods and Applications, 4, 077 (2008).

\bibitem{AdS1999} V. E. Adler, S. Ya. Startsev, “Discrete analogues of the Liouville equation”, Theoret. and Math. Phys., 121:2 (1999), 1484--1495. 

\bibitem{ShamsutdinovHab2009} M. A. Shamsutdinov, I. T. Habibullin, A. T. Kharisov, A. P. Tankeyev, “Dynamics of magnetic kinks in exchange-coupled ferromagnetic layers”, The Physics of Metals and Metallography, 108:4 (2009), 327–340.

\bibitem{Shamsutdinov2009} M. A. Shamsutdinov, I. Yu. Lomakina, V. N. Nazarov, A. T. Kharisov, D. M. Shamsutdinov,  “Ferro and antiferromagnetodynamics”, Nonlinear oscillations, waves and solitons. - M.: Nauka, 2009, 456 p.- ISBN 978-5-02-037454-6.

\bibitem{Kunin75} I. A. Kunin, “Elastic Media with Microstructure I”, Springer-Verlag Berlin Heidelberg New York 1982 Springer Series in Solid-State Sciences Editors: M. Cardona P. Fulde H.-J. Queisser.

\bibitem{AeroBulygin} E.L. Aero, A.N. Bulygin,  “The nonlinear theory of localized waves in a complex crystalline lattice as a discrete continual model” Computational Continuum Mechanics, 2008, Vol. 1 No. 1 (2008) pp 14-30.

\bibitem{Hab95} I. T. Habibullin, “Boundary conditions for integrable chains”, Physics Letters A, 207:5 (1995), 263–268 ,

\bibitem{FNR} E. V. Ferapontov, V. S. Novikov, I. Roustemoglou, “On the classification of discrete Hirota-type equations in 3D” Int. Math. Res. Not. IMRN,  \textbf{2015}:13 (2015), 4933--4974.

\bibitem{FHKN} E. V. Ferapontov, I. T. Habibullin, M. N. Kuznetsova, V. S. Novikov, “On a class of 2D integrable lattice equations”, J. Math. Phys., \textbf{61}:7 (2020), 073505.  

\bibitem{Hab13} I. Habibullin, “Characteristic Lie rings, finitely-generated modules and integrability conditions for (2+ 1)-dimensional lattices”, Physica Scripta, 87:6 (2013), 065005 , IOP Publishing 

\bibitem{HP2017}  I. Habibullin, M. Poptsova, “Classification of a Subclass of Two-Dimensional Lattices via Characteristic Lie Rings”, SIGMA, \textbf{13} (2017), 26 pp. 

\bibitem{HabKh21} I. T. Habibullin, A. R. Khakimova, “Characteristic Lie Algebras of Integrable Differential-Difference Equations in 3D”, J. Phys. A: Math. Theor., \textbf{54}:29 (2021), 295202.

\bibitem{KuzHabKh23} M. N. Kuznetsova, I. T. Habibullin, A. R. Khakimova, “On the problem of classifying integrable chain with three independent variables”,  Theoret. and Math. Phys., 215:2 (2023), 242–268

\bibitem{Smirnov15} S. V. Smirnov, “Semidiscrete Toda lattices”, Theoret. and Math. Phys., 172:3 (2012), 1217–1231

\bibitem{ZhiberK21} A. V. Zhiber, M. N. Kuznetsova, “Integrals and characteristic Lie rings of semi-discrete systems of equations”, Ufa Math. J., 13:2 (2021), 22–32

\bibitem{HabKh22} I. T. Habibullin, A. R. Khakimova, “Integrals and characteristic algebras for systems of discrete equations on a quadrilateral graph”, Theoret. and Math. Phys., 213:2 (2022), 1589–1612

\end{thebibliography}
\end{document}